\documentstyle[preprint,prc,aps]{revtex}
\begin{document}
\draft
\title{\bf Time Invariance Violating Nuclear Electric Octupole
Moments}
\author{V. V. Flambaum$^{1,2}$, D. W. Murray$^1$,
and S. R. Orton$^1$}
\address{$^1$School of Physics, University of New South Wales,
Sydney, 2052, Australia\\
$^2$  ITAMP, Harvard University and the Smithsonian Astrophysical
Observatory,
60 Garden Street, Cambridge, Massachusetts 02138}
\maketitle
\begin{abstract}
The existence of a nuclear electric octupole moment (EOM)
requires both parity and time invariance violation.
The EOMs of odd $Z$ nuclei that are
induced by a particular T- and P-odd interaction are
calculated. We compare such octupole moments with the collective
EOMs
that can occur in nuclei having a static octupole
deformation. A nuclear EOM can induce a parity and
time invariance violating atomic electric dipole moment, and the
magnitude of this effect is calculated. The contribution of
a nuclear EOM to such a dipole moment is
found, in most cases, to be smaller than that of
other mechanisms of atomic electric dipole moment production.
\end{abstract}
\vspace{5mm}
\pacs{PACS numbers: 21.10.Ky, 11.30.Er, 32.80.Ys, 32.10.Dk}
\vspace{10mm}

\section{Introduction}
At present, time invariance violation has only been observed
indirectly,
in the CP-violating decay
of neutral K-mesons \cite{CPv}. Parity and
time invariance violating nuclear and
atomic multipole moments, such
as the magnetic monopole, electric dipole, magnetic quadrupole,
and electric octupole moments, are interesting
because, if discovered,
they would provide further proof of time invariance violation.
Even the limits on these moments provide important tests of
different models of CP-violation.
Parity and time invariance violating nuclear moments induced by
T- and P-odd nuclear forces have been discussed, e.g., in
Refs. \cite{Fein77,Cov83,Hax83,SFK84,FKS86,Khripl}.

In this paper we consider the electric octupole
moments of nuclei. We calculate the EOMs of odd $Z$ nuclei that
are induced by a T- and P-odd interaction
between the unpaired proton
and the nuclear core. Note that nuclei with unpaired
neutrons can
have an EOM of comparable magnitude due to the polarization of the
nuclear core by the T-, P-odd field of the external neutron. A
similar mechanism for the electric dipole and Schiff moments was
considered in Ref. \cite{FKS86}.

We give values of the collective electric
octupole moments of nuclei having a static octupole 
deformation, using the calculations done
in Refs. \cite{AFS96,AFS2}. We also present a
calculation of the atomic electric dipole moment (EDM) 
that would be induced by a nuclear EOM. Finally, we discuss
a possible enhancement mechanism for
magnetic quadrupole moments (MQMs) in nuclei with
octupole deformation.

In the appendix we present simple estimates of
the relative sizes of the contributions
of various nuclear moments to the atomic EDM. The magnitude of
the nuclear EOM is comparable to that of the Schiff moment.
However, the contribution of the EOM to the atomic EDM is
smaller than the contribution of the Schiff moment (and the
magnetic quadrupole moment) since the EOM interacts with
higher angular momentum electron states, whose wave functions
are suppressed near the nucleus. 

\section{The Hamiltonian of the T-, P-odd nucleon-nucleus
interaction and the resulting nucleon wave function}
\label{shptnw}
For a heavy nucleus, the T- and P-odd interaction between a
nonrelativistic unpaired nucleon and the nuclear core can be 
described
by the following effective Hamiltonian (see, e.g.,
Refs. \cite{Henley,SFK84,Khripl}):
\begin{equation}
H_{TP} = \eta \frac{G}{2 \sqrt{2} m} \bbox{\sigma} \cdot
\bbox{\nabla} \rho,
\label{ehpta}
\end{equation}
where $\bbox{\sigma}$ is twice the spin operator for this nucleon,
$\rho$ is the density of the nuclear core,
$G = 1.0 \times 10^{-5}/{m_p}^2$ is the Fermi constant,
$m$ is the mass of the nucleon and $\eta$ is a dimensionless
constant that describes the strength of the interaction.
In this paper we
deal with odd $Z$ nuclei, and so the nucleon involved is the
unpaired proton.

Let $U$ be the strong nuclear potential of the core that the 
unpaired proton moves in. The range of
the strong nucleon-nucleon interaction
is small. This means that the potential $U({\bf r})$ and 
the nuclear density $\rho ({\bf r})$ will be similar in shape.
In fact, we assume
that they are approximately proportional: $U({\bf r})/U({\bf 0})
\approx \rho({\bf r})/\rho({\bf 0})$. This allows us to rewrite
Eq.\ (\ref{ehpta}) in a form that makes it
easy to find the perturbed
wave function due to $H_{TP}$ (we use the method of
Ref. \cite{SFK84}). We obtain
\begin{equation}
H_{TP} \approx \xi \bbox{\sigma} \cdot \bbox{\nabla} U,
\label{ehptb}
\end{equation}
where
\begin{equation}
\xi = \eta \frac{G}{2 \sqrt{2} m_p} \frac{\rho ({\bf 0})}
{U ({\bf 0})} = -2 \times 10^{-21} \eta \mbox{ cm}.
\label{ehptc}
\end{equation}
The total potential that the unpaired proton experiences is
\begin{equation}
\widetilde{U} = U + H_{TP} \approx U({\bf r})
+ \xi \bbox{\sigma} \cdot \bbox{\nabla} U \approx
U({\bf r} + \xi \bbox{\sigma}).
\label{ehptd}
\end{equation}
As a result, if $\psi ({\bf r})$ is the proton's wave function
when only
$U({\bf r})$ is present, the perturbed wave function will be
\begin{equation}
\widetilde{\psi} ({\bf r}) \approx \psi ({\bf r} + \xi
\bbox{\sigma})
\approx \psi({\bf r}) + \xi \bbox{\sigma} \cdot \bbox{\nabla}
\psi({\bf r}).
\label{ehpte}
\end{equation}

\section{The electric octupole moment of a nucleus with
an unpaired proton}
\label{sneom}
In this section we calculate the electric octupole moment
(EOM) of a
nucleus with an unpaired proton using
the perturbed wave function obtained above.
The electric octupole moment can be written as
[see Eq. (\ref{eneo8caa})]
\begin{equation}
O_{ijk} = \langle \widetilde{\psi} | \hat{O}_{ijk} |
\widetilde{\psi} \rangle
= e \int \widetilde{\psi}^{\dagger}({\bf r})
[r_i r_j r_k
- {\textstyle \frac{1}{5}} r^2 (r_i \delta_{jk} + r_j \delta_{ik}
+ r_k \delta_{ij})] \widetilde{\psi} ({\bf r}) \, d^3 r.
\label{eaa}
\end{equation}
Note that we use $e > 0$.
Substituting $\widetilde{\psi} ({\bf r}) =
\psi ({\bf r}) + \xi \sigma_m 
\frac{\partial \psi}{\partial r_m}$ (from Eq.\ (\ref{ehpte}))
into the above integral and expanding gives
\begin{equation}
O_{ijk} = 2 e \xi \int \frac{\partial \psi^{\dagger}}{\partial r_m}
\sigma_m [r_i r_j r_k - 
{\textstyle \frac{1}{5}} r^2 (r_i \delta_{jk} + r_j \delta_{ik}
+ r_k \delta_{ij})] \psi \, d^3 r.
\label{eneom1a}
\end{equation}
Notice that we have discarded the $\xi^2$ term
and that the term not
containing $\xi$ vanishes, as it is an integral of an odd function
of ${\bf r}$. Applying integration by parts to the above
integral gives 
\begin{eqnarray}
O_{ijk} & =  - {\textstyle \frac{1}{5}} e \xi \langle \psi
| & 
5 (\hat{r}_i \hat{r}_j \hat{\sigma}_k +
\hat{r}_i \hat{r}_k \hat{\sigma}_j +
\hat{r}_j \hat{r}_k \hat{\sigma}_i)
-{\hat{r}}^2 (\hat{\sigma}_i \delta_{jk} +
\hat{\sigma}_j \delta_{ik}
+ \hat{\sigma}_k \delta_{ij}) \nonumber \\
&& - 2 \hat{r}_m \hat{\sigma}_m
(\hat{r}_i \delta_{jk} + \hat{r}_j \delta_{ik} + \hat{r}_k
\delta_{ij})
|
\psi \rangle.
\label{eac}
\end{eqnarray}

We can also write the octupole moment tensor in another form.
Since it is a symmetric, irreducible (traceless) third rank tensor
and the nuclear angular momentum ${\bf I}$ is the only 
quantity which defines
a direction in the system, the EOM tensor must be in the form
of the most general symmetric, irreducible third rank tensor that
can be formed from the components of $\hat{\bf I}$. That is
\begin{equation}
O_{ijk} = \langle \widetilde{\psi} | \hat{O}_{ijk} |
\widetilde{\psi} \rangle ,
\label{ead}
\end{equation}
where
\begin{eqnarray}
\hat{O}_{ijk} & = A [ & \hat{I}_i \hat{I}_j \hat{I}_k +
\hat{I}_j \hat{I}_k \hat{I}_i + \hat{I}_k \hat{I}_i \hat{I}_j
+ \hat{I}_k \hat{I}_j \hat{I}_i + \hat{I}_j \hat{I}_i \hat{I}_k
+ \hat{I}_i \hat{I}_k \hat{I}_j \nonumber \\
&& - {\textstyle \frac{6I(I+1)-2}{5}}
(\hat{I}_i \delta_{jk} + \hat{I}_j \delta_{ik}
+ \hat{I}_k \delta_{ij} )]
\label{eae}
\end{eqnarray}
and $A$ is some constant. The factor of $-[6I(I+1)-2]/5$ follows
from the requirement
of tracelessness ($O_{iij}=O_{iji}=O_{jii}=0$).
The quantity which is usually referred to as the octupole moment
is ${\cal O}$, which is the $O_{zzz}$ component for the nuclear
state having
angular momentum projection $I_z = I$. This is the quantity
which we will calculate. We can write $A$ in terms of ${\cal O}$
using Eqs.\ (\ref{ead}) and (\ref{eae}), with $i=j=k=z$. This gives
\begin{eqnarray}
\hat{O}_{ijk} & =
\frac{5 {\cal O}}{6I(I-1)(2I-1)}
[ & \hat{I}_i \hat{I}_j \hat{I}_k +
\hat{I}_j \hat{I}_k \hat{I}_i + \hat{I}_k \hat{I}_i \hat{I}_j
+ \hat{I}_k \hat{I}_j \hat{I}_i + \hat{I}_j \hat{I}_i \hat{I}_k
+ \hat{I}_i \hat{I}_k \hat{I}_j \nonumber \\
&& - {\textstyle \frac{6I(I+1)-2}{5}}
(\hat{I}_i \delta_{jk} + \hat{I}_j \delta_{ik}
+ \hat{I}_k \delta_{ij} )].
\label{eaf}
\end{eqnarray}
We now have two expressions for $O_{ijk}$: Eq.\ (\ref{eac}) and
Eq.\ (\ref{eaf}) (via Eq.\ (\ref{ead})). We calculate ${\cal O}$ by
operating on both of these equations with $\hat{I}_i \hat{I}_j$ on
the left and $\hat{I}_k$ on the right and equating the results.

To evaluate the results of these operations, the following
commutation
relations are required: $[\hat{r}_i,\hat{\sigma}_j] = 0$,
$[\hat{I}_i,\hat{r}_j] = i \varepsilon_{ijk} \hat{r}_k$, and
$[\hat{I}_i,\hat{\sigma}_j] = i \varepsilon_{ijk} \hat{\sigma}_k$.
(Note that $\hat{I}_i = \hat{l}_i + \hat{\sigma}_i / 2$.)
Other useful relations are
$[\hat{I}_i,\hat{I}_j \hat{r}_j] = [\hat{I}_i,\hat{I}_j
\hat{\sigma}_j]
= [\hat{I}_i,\hat{\sigma}_j \hat{r}_j] = 0$ and
$\varepsilon_{ijk} A_i A_j = {\textstyle \frac{1}{2}}
\varepsilon_{ijk} A_i A_j
- {\textstyle \frac{1}{2}} \varepsilon_{ijk} A_j A_i
= {\textstyle \frac{1}{2}} \varepsilon_{ijk} [A_i,A_j]$,
for an operator $A_i$.
The commutation relations
are used to rearrange the operators so that pairs
with the same indices are adjacent.

Applying the operators to
Eq. (\ref{eac}) gives
\begin{eqnarray}
\langle \widetilde{\psi} | \hat{I}_i \hat{I}_j \hat{O}_{ijk}
\hat{I}_k | \widetilde{\psi} \rangle
& = -{\textstyle \frac{1}{5}} e \xi \langle \psi | &
5 \hat{I}_i \hat{I}_j (\hat{r}_i \hat{r}_j \hat{\sigma}_k +
\hat{r}_i \hat{r}_k \hat{\sigma}_j + \hat{r}_j \hat{r}_k
\hat{\sigma}_i)
\hat{I}_k
- \hat{I}_i \hat{I}_j \hat{r}^2 (\hat{\sigma}_i \delta_{jk} +
\hat{\sigma}_j \delta_{ik}
+ \hat{\sigma}_k \delta_{ij}) \hat{I}_k \nonumber \\
&& - 2 \hat{I}_i \hat{I}_j \hat{r}_m \hat{\sigma}_m 
(\hat{r}_i \delta_{jk} + \hat{r}_j \delta_{ik} + \hat{r}_k
\delta_{ij})
\hat{I}_k | \psi \rangle \nonumber \\
& = -{\textstyle \frac{1}{5}} e \xi \langle r^2 \rangle \{ &
\langle I,I_z,l | (\hat{\bbox{\sigma}} \cdot {\hat{\bf n}})
[ {\textstyle \frac{5}{4}} (\hat{\bf I} \cdot \hat{\bbox{\sigma}})
- \hat{I}^2 ] (\hat{\bbox{\sigma}} \cdot {\hat{\bf n}})
| I,I_z,l \rangle \nonumber \\
&&
+ \langle I,I_z,l | {\textstyle \frac{7}{2}} (\hat{\bf I} \cdot
\hat{\bbox{\sigma}})
- 3 \hat{I}^2 (\hat{\bf I} \cdot \hat{\bbox{\sigma}}) + 1
- 2 \hat{I}^2 | I,I_z,l \rangle \},
\label{eal}
\end{eqnarray}
where $\hat{\bf n} = \hat{\bf r} / r$.
Here the facts that
$\hat{\bf I} \cdot \hat{\bf n} = (\hat{\bf l} +
\hat{\bbox{\sigma}}/2)
\cdot \hat{\bf n} = (\hat{\bbox{\sigma}} \cdot \hat{\bf n})/2$
and $(\hat{\bbox{\sigma}} \cdot \hat{\bf n})^2 = 1$ \cite{QMNRT}
were used. 
The radial and angular parts of the
operators have been separated in Eq.\ (\ref{eal});
$\langle r^2 \rangle$
is the expectation value of $r^2$ and $| I,I_z,l \rangle$ is the
angular part of $| \psi \rangle$.
The second matrix element in Eq.\ (\ref{eal}) is
\begin{equation}
\langle I,I_z,l | {\textstyle \frac{7}{2}}
(\hat{\bf I} \cdot \hat{\bbox{\sigma}})
- 3 \hat{I}^2 (\hat{\bf I} \cdot \hat{\bbox{\sigma}}) + 1
- 2 \hat{I}^2 | I,I_z,l \rangle
= [{\textstyle \frac{7}{2}} - 3I(I+1)]
({\textstyle \frac{1}{2}} - \kappa) + 1 -2I(I+1),
\label{eam}
\end{equation}
where we have used the fact that
\begin{eqnarray}
(\hat{\bf I} \cdot \hat{\bbox{\sigma}}) | I,I_z,l \rangle & =
& [I(I+1) - l(l+1) + 3/4] | I,I_z,l \rangle \nonumber \\
& = & (1/2 - \kappa) | I,I_z,l \rangle,
\label{ean}
\end{eqnarray}
with
\begin{equation}
\kappa = (I + 1/2) (-1)^{I+1/2-l}.
\label{eao}
\end{equation}
To evaluate the first matrix element in Eq.\ (\ref{eal}), we use
the following identity:
\begin{equation}
(\hat{\bbox{\sigma}} \cdot \hat{{\bf n}}) | I,I_z,l \rangle
= - | I,I_z,\widetilde{l} \rangle,
\label{eap}
\end{equation}
where $\widetilde{l} = 2 I - l$.
We then have
\begin{eqnarray}
\langle I,I_z,l | (\hat{\bbox{\sigma}} \cdot {\hat{\bf n}})
[ {\textstyle \frac{5}{4}} (\hat{\bf I} \cdot \hat{\bbox{\sigma}})
- \hat{I}^2 ] (\hat{\bbox{\sigma}} \cdot {\hat{\bf n}})
| I,I_z,l \rangle & = &
\langle I,I_z,\widetilde{l} | {\textstyle \frac{5}{4}} (\hat{\bf I}
\cdot \hat{\bbox{\sigma}})
- \hat{I}^2 | I,I_z,\widetilde{l} \rangle \nonumber \\
& = & {\textstyle \frac{5}{4}} ({\textstyle \frac{1}{2}}
+ \kappa) - I(I+1).
\label{eaq}
\end{eqnarray}
(Note that $(\hat{\bf I} \cdot \hat{\bbox{\sigma}})
| I,I_z,\widetilde{l} \rangle = (1/2 + \kappa)
| I,I_z,\widetilde{l} \rangle$.)
Using these results in Eq.\ (\ref{eal}) gives
\begin{equation}
\langle \widetilde{\psi} | \hat{I}_i \hat{I}_j \hat{O}_{ijk}
\hat{I}_k
| \widetilde{\psi} \rangle = -{\textstyle \frac{3}{5}} e \xi
\langle r^2 \rangle
(\kappa - {\textstyle \frac{3}{2}})
(I-{\textstyle \frac{1}{2}})(I+{\textstyle \frac{3}{2}}).
\label{ear}
\end{equation}
Now we find another form of the left hand side of the above
equation by operating on Eq.\ (\ref{eaf}). We have
\begin{eqnarray}
\langle \widetilde{\psi} | \hat{I}_i
\hat{I}_j \hat{O}_{ijk} \hat{I}_k
| \widetilde{\psi} \rangle
& =
\frac{5 {\cal O}}{6I(I-1)(2I-1)}
\langle \widetilde{\psi} |
& \hat{I}_i \hat{I}_j (\hat{I}_i \hat{I}_j \hat{I}_k +
\hat{I}_j \hat{I}_k \hat{I}_i + \hat{I}_k \hat{I}_i \hat{I}_j
+ \hat{I}_k \hat{I}_j \hat{I}_i + \hat{I}_j \hat{I}_i \hat{I}_k
+ \hat{I}_i \hat{I}_k \hat{I}_j) \hat{I}_k \nonumber \\
&& - {\textstyle \frac{6I(I+1)-2}{5}} \hat{I}_i \hat{I}_j 
(\hat{I}_i \delta_{jk} + \hat{I}_j \delta_{ik}
+ \hat{I}_k \delta_{ij} ) \hat{I}_k | \widetilde{\psi} \rangle.
\label{eas}
\end{eqnarray}
This can be evaluated using the equation
$[\hat{I}_i,\hat{I}_j] = i \varepsilon_{ijk} \hat{I}_k$ to
bring operators with the same indices together
and $\varepsilon_{ijk} \hat{I}_i \hat{I}_j =
\frac{i}{2}
\varepsilon_{ijk} \varepsilon_{ijp} \hat{I}_p = i \hat{I}_k$.
The result is
\begin{equation}
\langle \widetilde{\psi} |
\hat{I}_i \hat{I}_j \hat{O}_{ijk} \hat{I}_k
| \widetilde{\psi} \rangle = {\cal O} (I+1)(I+3/2)(I+2).
\label{eat}
\end{equation}
Equating this result with Eq.\ (\ref{ear}) gives the following
result for the octupole moment:
\begin{equation}
{\cal O}_{\rm sing} = \frac{-3(\kappa-3/2)(I-1/2)}
{5(I+1)(I+2)} \langle r^2 \rangle e \xi,
\label{eau}
\end{equation}
(The subscript refers to the fact that this octupole moment is
due to a single particle, as opposed to a collective octupole
moment.)
The expectation value for $r^2$ can be
approximated as \cite{SFK84}
\begin{equation}
\langle r^2 \rangle
\approx {\textstyle \frac{3}{5}} {r_0}^2 A^{2/3},
\label{eaw}
\end{equation}
to an accuracy of about 10\%,
where $r_0 = 1.1 \mbox{ fm}$ and $A$ is the mass number of
the nucleus. Using the value of $r_0$ and Eqs.\ (\ref{ehptc}) and
(\ref{eao}) gives
\begin{equation}
{\cal O_{\rm sing}} \approx 8.7 \times 10^{-9} A^{2/3}
\eta e ({\rm fm})^3 \times
\left\{
\begin{array}{ll}
\frac{-(I-1/2)}{I+1} & \mbox{for $I=l+1/2$} \\
\frac{(I-1/2)(I-1)}{(I+1)(I+2)} & \mbox{for $I=l-1/2$}
\end{array} \right..
\label{eaw2}
\end{equation}
Observe that ${\cal O}_{\rm sing}=0$ for $I=1/2$. This
is to be expected because
$\hat{O}_{ijk}$ is a third rank tensor, and so applying the
triangle rule for the addition of angular momenta
to Eq.\ ({\ref{ead})
gives the result that nuclei having angular momentum less
than $3/2$ cannot have an octupole moment.
Values of ${\cal O}_{\rm sing}$ for various nuclei
are given in table \ref{tsingp}, in terms of the parameter $\eta$.
  
\section{Collective electric octupole moments in nuclei with static
octupole deformation}
\label{sceom}
In Refs.\ \cite{AFS96,AFS2}, a mechanism was suggested by which
parity and time invariance
violating interactions can produce collective T- and P-odd
multipole moments in 
even-odd nuclei having a static octupole deformation
[i.e. electric octupole moments in their intrinsic
(or body-fixed) reference frames].
Such a deformation has been shown to exist for nuclei in the
Ra--Th and Ba--Sm regions (for a review see, e.g., \cite{Ahmad93}).
A similar
mechanism, for the enhancement of the intrinsic electron EDM and
other T-, P-odd interactions in polar molecules, was suggested in
Ref. \cite{SF1978}.
Below, we explain the mechanism by which a collective nuclear
EOM can be produced and provide values of this EOM for various
nuclei.

\subsection{Parity and time invariance and octupole moments in
the laboratory frame}
An electric octupole moment can exist in the nucleus's intrinsic
frame without parity or time invariance violation. Yet if parity
and time invariance hold, the expectation value of the octupole
moment in the laboratory reference frame will be zero.

Consider
$|I M K \rangle$ and $|I M -K \rangle$, which are two almost
degenerate states of the nucleus in the laboratory
frame. $I$ is the angular momentum of the nucleus,
$M$ is its projection onto the $z$-axis, and $K$
is its projection onto
the axis of symmetry of the deformed nuclear core
(the $z^{\prime}$-axis).
(${\bf I}$ is the sum of the unpaired nucleon's angular
momentum, ${\bf j}$ and the nuclear core's orbital angular
momentum, ${\bf R}$.)
These states can be written in terms of intrinsic states as
\begin{equation}
|I M \pm K \rangle =
\sqrt{\textstyle \frac{2I+1}{4 \pi}}
 D^I_{M \pm K} (\phi,\theta,0) \psi_{\pm K}
({\bf r}^\prime) \chi_{\rm core},
\label{ecn21a}
\end{equation}
where $D^I_{M \pm K} (\phi,\theta,0)$ is a Wigner $D$-function
(see, e.g., \cite{Varshalovich,RelQuant}),
$\chi_{\rm core}$ is the wave function of the nuclear core
in the intrinsic frame, and
$\psi_{\pm K} ({\bf r}^\prime)$ is the wave function of the unpaired
nucleon in the intrinsic frame, with
a $z^{\prime}$ angular momentum projection of $\pm K$. (In the
intrinsic frame, the nuclear axis plays the role of the usual
``$z$-axis'' in Quantum Mechanics.) Note that ${\bf j}$ and
${\bf I}$ have the same $z^{\prime}$ projection.

$|I M K \rangle$ and $|I M -K \rangle$ do not
have good parity, as $K$
changes sign under a parity transformation. However, the following
states do, and they form a parity doublet:
\begin{equation}
\psi^{\pm} = \frac{1}{\sqrt{2}} ( |I M K \rangle
\pm |I M -K \rangle).
\label{eceom1a}
\end{equation}
For these good parity states 
$\langle \psi^{\pm} | {\bf I} \cdot {\bf n} | \psi^{\pm}
\rangle = 0$
because $K$ and $-K$ have equal probabilities
and this means that there is no
average orientation of the nuclear axis in the laboratory frame
($\langle \psi^{\pm} | {\bf n} | \psi^{\pm} \rangle = {\bf 0}$).
This is a consequence of time invariance and parity
conservation since
the correlation ${\bf I} \cdot {\bf n}$ is T-, P-odd. As a result
of $\langle \psi^{\pm} | {\bf n} | \psi^{\pm} \rangle = {\bf 0}$,
the mean value of the octupole moment
(whose orientation is determined by the direction of the
nuclear axis)
is zero in the laboratory frame.  

Now, a T- and P-odd interaction, $H_{TP}$
will mix the members of the parity doublet ($\psi^+$ and $\psi^-$).
The admixed wave function of the predominantly positive parity
member of the doublet will be $\psi = \psi^+ + \alpha \psi^-$ or
\begin{equation}
\psi = \frac{1}{\sqrt{2}} [(1+\alpha) |I M K \rangle
+ (1-\alpha) |I M -K \rangle],
\label{eceom2c}
\end{equation}
where $\alpha$ is a mixing coefficient:
\begin{equation}
\alpha = \frac{\langle \psi^- | H_{TP} |\psi^+ \rangle}{E_+ - E_-}.
\label{enad32}
\end{equation}
$E_+ - E_-$ is the energy splitting between the members of the
parity doublet.
The interaction $H_{TP}$ is given by Eq.\ (\ref{ehpta}).
(A similar expression can be obtained for the predominantly
negative parity member of the doublet.)
This mixing yields, on average,
an orientation of the nuclear axis along the direction of the
angular momentum:
\begin{equation}
\langle \psi | {\bf I} \cdot {\bf n} | \psi \rangle
= \langle \psi | \hat{K} | \psi \rangle
= 2 \alpha K,
\label{eceom2d}
\end{equation}
and this means that the octupole moment need no longer vanish in
the laboratory frame.

\subsection{The magnitude of the collective octupole moment}
The collective EOM in the laboratory frame
was derived in \cite{AFS96,AFS2},
with the following result:
\begin{equation}
{\cal O}_{\rm coll} \approx \frac{4}{5}
\frac{I (I-1) (I - 1/2)}{(I+1)
(I+2) (I + 3/2)}
\alpha O_{3,{\rm intr}}.
\label{eceom1}
\end{equation}
Once again, observe that ${\cal O}_{\rm coll} = 0$ for $I=1/2$.
$O_{3,{\rm intr}}$ refers to the octupole moment in the
intrinsic frame ($ O_{zzz} \equiv \frac{2}{5} O_3$)
and is given by \cite{BohrMott,LeanderChen}:
\begin{equation}
O_{3,{\rm intr}} = e Z {R_0}^3 \frac{3}{2 \sqrt{7 \pi}}
(\beta_3 + \frac{2}{3} \sqrt{\frac{5}{\pi}} \beta_2 \beta_3 + 
\frac{15}{11 \sqrt{\pi}} \beta_3 \beta_4 + \ldots),
\label{eceom2}
\end{equation}
where $R_0 = r_0 A^{1/3}$ ($r_0 = 1.1 \mbox{ fm}$). 
$\beta_2$, $\beta_3$, and $\beta_4$ are parameters
that describe the nuclear
deformation; the surface of a deformed nucleus is
\begin{equation}
R = R_0 [1 + \sum_{l=1}^{\infty} \beta_l Y_{l0} (\theta,\phi) ].
\label{eceom3}
\end{equation}

We will first present an order of magnitude estimate of
$\alpha$.
$\hat{K} = {\bf I} \cdot {\bf n}$ and $H_{TP}$ are both
T-, P-odd pseudoscalars. Therefore,
$\langle \psi_{+K} | H_{TP} | \psi_{+K} \rangle \propto K$ and
so $\langle \psi_{-K} | H_{TP} | \psi_{-K} \rangle
= -\langle \psi_{+K} | H_{TP} | \psi_{+K} \rangle$ (this fact can
be easily supported by model calculations). Using this fact
and Eqs.\ (\ref{ecn21a}) and (\ref{eceom1a}) we get
$\langle \psi^- | H_{TP} | \psi^+ \rangle
= \langle \psi_{+K} | H_{TP} | \psi_{+K} \rangle$. If
$\psi_{+K}$ were a good parity state this matrix element would
be zero. However, due to the perturbation caused by the static
octupole deformation of the nucleus ($V_3$), it is a combination
of the
opposite parity spherical orbitals $\psi_{1,+K}$
and $\psi_{2,+K}$ (e.g., $p_{3/2}$ and $d_{3/2}$):
\begin{equation}
\psi_{+K} = \psi_{1,+K} + \gamma \psi_{2,+K},
\label{eae11a}
\end{equation}
\begin{equation}
\gamma = \frac{\langle \psi_{2,+K} | V_3 | \psi_{1,+K}
\rangle}{E_1 - E_2},
\label{eae11b}
\end{equation}
\begin{eqnarray}
\psi_{1,+K} & = & R_1(r^{\prime}) \Omega_{j,l,+K} (\theta^{\prime},
\phi^{\prime}), \nonumber \\
\psi_{2,+K} & = & R_2(r^{\prime}) \Omega_{j,\widetilde{l},+K}
(\theta^{\prime},\phi^{\prime})
= - R_2(r^{\prime}) (\bbox{\sigma}
\cdot {\bf n}^{\prime}) \Omega_{j,l,+K}
(\theta^{\prime},\phi^{\prime}),
\label{eae11c}
\end{eqnarray}
where $\widetilde{l} = 2j-l$.
(Of course there will also be an admixture of
other opposite parity states
having different values of $j$. We neglect these states
for simplicity.)
Therefore, we have
\begin{equation}
\alpha = \frac{\langle \psi_{+K}|H_{TP}|\psi_{+K} \rangle}{E_+-E_-}
= 2 \gamma \frac{\langle
\psi_{1,+K} | H_{TP} | \psi_{2,+K} \rangle}{E_+
- E_-}.
\label{ena42}
\end{equation}

To estimate $\gamma$ we must first derive the form of $V_3$.
As in Sec. \ref{shptnw}, let $U$ be
the strong nuclear potential that
the unpaired nucleon moves in. $V_3$
is the difference between the potentials with the octupole
deformation present, $U_{\rm pres}$ and
absent, $U_{\rm abs}$. Once again we have
$U({\bf r}^{\prime}) \approx U({\bf 0}) \rho({\bf r}^{\prime}) /
\rho ({\bf 0})$. We also make the approximation that
$\rho({\bf r}^{\prime}) / \rho ({\bf 0}) \approx
\theta (r^{\prime} - R)$, where $R$ is the nuclear radius and
$\theta(x) = 1$ for $x<0$ and $\theta(x) = 0$ for $x>0$. For an
octupole deformed nucleus we have $R = R_0 (1 + \beta_3 Y_{30})$
and so
\begin{equation}
V_3 = U_{\rm pres} - U_{\rm abs}
\approx U({\bf 0}) [\theta(r^{\prime}- R_0 - \beta_3 Y_{30} R_0)
- \theta(r^{\prime} - R_0)]
\approx U({\bf 0}) R_0 \beta_3 \delta(r^{\prime}-R_0) Y_{30},
\label{eae21a}
\end{equation}
where we have expanded $\theta$ in a Taylor series, using
$\frac{d \theta}{dx} = -\delta(x)$.
Using Eq.\ (\ref{eae11b}) we then
have
\begin{equation}
| \gamma | \approx \left| U({\bf 0})
\beta_3 R_1(R_0) R_2(R_0) {R_0}^3
\int \Omega_2^{\dagger} Y_{30} \Omega_1 \, d \Omega^{\prime}
(E_1 - E_2)^{-1} \right|
\sim \beta_3,
\label{eae21b}
\end{equation}
where we have used $R_1 (R_0) R_2 (R_0) \approx 1.4 / {R_0}^3$
\cite{BohrMott}, $| U({\bf 0}) | \approx 50 \mbox{ MeV}$,
$|E_1-E_2| \approx 5 \mbox{ MeV}$, and
$\int \Omega_2^{\dagger} Y_{30} \Omega_1 \, d \Omega^{\prime} \sim
0.05$.

Finally, we must estimate
the matrix element between the spherical orbitals,
$\langle \psi_{1,+K} | H_{TP} | \psi_{2,+K} \rangle$.
Using Eq.\ (\ref{ehpta}) for
the form of $H_{TP}$ and $\rho(r^{\prime})
= \theta(r^{\prime}-R_0) / ({\textstyle \frac{4}{3}} \pi {r_0}^3)$
we get
\begin{equation}
H_{TP} = -\eta \frac{3G}{8 \pi \sqrt{2} m {r_0}^3} (\bbox{\sigma}
\cdot {\bf n}^{\prime}) \delta(r^{\prime}-R_0).
\label{eae31c}
\end{equation}
Using Eq.\ (\ref{eae11c}) and $(\bbox{\sigma}
\cdot {\bf n}^{\prime})^2 = 1$
gives
\begin{equation}
\langle \psi_{1,+K} | H_{TP} | \psi_{2,+K} \rangle
= \eta \frac{3G}{8 \pi \sqrt{2} m {r_0}^3}
R_1(R_0) R_2(R_0) {R_0}^2
\approx \frac{\eta}{A^{1/3}} 1 \mbox{ eV}.
\label{eae31d}
\end{equation}

Using $|E_+ - E_-| \sim 50 \mbox{ keV}$
(see, e.g., \cite{AFS96,AFS2}),
$\beta_3 \approx 0.1$ (see, e.g., \cite{Cwiok91}),
and Eqs.\ (\ref{ena42}), (\ref{eae21b}), and (\ref{eae31d})
gives (for $A \approx 225$)
$|\alpha| \sim 2 \beta_3 A^{-1/3} \eta
\, {\rm eV} / |E_+ - E_-|
\sim 7 \times 10^{-7} \eta$.
This provides the following estimate for the collective EOM:
\begin{equation}
| {\cal O}_{\rm coll} |
\sim 0.05 e {\beta_3}^2 Z A^{2/3} {r_0}^3 \eta
\, {\rm eV} / |E_+ - E_-|
\sim 4 \times 10^{-5} \eta e ({\rm fm})^3.
\label{eeome30}
\end{equation}
We see that the collective EOM is two orders of magnitude
larger than the EOM due to unpaired protons.

We can do a more accurate calculation of the EOM
using Refs. \cite{AFS96,AFS2},
where the mixing coefficients, $\alpha$ for various
nuclei were calculated.
We use the values from \cite{AFS2} that were calculated using
the Woods-Saxon potential.
We took the values of the $\beta_i$ parameters from
\cite{Cwiok91} and the nuclei's angular momenta were taken
from \cite{Butler96,LeanderChen}.
The results are
shown in table \ref{tcoll} for various nuclei.

Note the large value of $^{229}$Pa's collective octupole moment.
This is due to its large value of $\alpha$, which is caused by
the small energy splitting between the members of
its parity doublet \cite{AFS96,AFS2}.
The possible existence of a static octupole deformation
in $^{229}$Pa was stated in \cite{Ahmad82}. However, more recent
papers cast doubt on the existence of such a
deformation (see, e.g., \cite{Grafen91,Levon94}).
Therefore, the result given for $^{229}$Pa must be
understood as being
conditional on it having a static octupole deformation.

\section{The atomic electric dipole moment induced
by a nuclear electric octupole moment}
\label{saedmi}
In this section we consider the electric dipole moment of an
atom that is induced by a nuclear electric octupole
moment.
The electric potential, $\phi ({\bf r})$ outside an arbitrary
charge distribution can be expanded in
terms of spherical harmonics as (see, e.g., \cite{Jackson})
\begin{equation}
\phi ({\bf r}) = \sum_{l=0}^{\infty} \sum_{m=-l}^l
\frac{4 \pi}{2l+1} q_{lm} \frac{Y_{lm} (\theta,\phi)}{r^{l+1}}.
\label{eedm1}
\end{equation}
Here we neglect that part of the potential that comes
from the screening of the nucleus's Coulomb field by
the atomic electrons as it
is not important in the octupole potential (see the
appendix).
The $q_{lm}$ are spherical electric multipole moments and they
can be written in terms of the charge distribution
$\rho_c ({\bf r}^{\prime})$ as follows \cite{Jackson}:
\begin{equation}
q_{lm} = \int Y_{lm}^* (\theta^\prime,\phi^\prime) 
{r^\prime}^l \rho_c ({\bf r}^{\prime}) \, d^3 r^\prime.
\label{eedm2}
\end{equation}
The octupole term of the above electric potential is
\begin{equation}
\phi^{(3)} ({\bf r}) = \frac{4 \pi}{7} q_{30} \frac{1}{r^4}
Y_{30} (\theta,\phi),
\label{eedm3}
\end{equation}
where we have only taken the $m=0$ term, as we will be
calculating the matrix element of the perturbation between
states with the same angular momentum projections. We can
write this in terms of the octupole moment by using the
relation $q_{30} = \int Y_{30} (\theta,\phi) r^3 \rho_c
({\bf r}) \, d^3 r
= \frac{5}{4} \sqrt{\frac{7}{\pi}} O_{zzz}$.
The potential energy, $U_{\rm oct}$ of an electron
(of charge $-e$) in
$\phi^{(3)} ({\bf r})$ will then be
\begin{equation}
U_{\rm oct} = -5 \sqrt{\frac{\pi}{7}} e {\cal O} \frac{1}{r^4} 
Y_{30} (\theta,\phi)
\label{eedm4}
\end{equation}
(for the nuclear $I_z = I$ state).

Using perturbation theory, the electric dipole moment of,
for example,
an atom with one electron above closed subshells induced
by $U_{\rm oct}$ can be written as
\begin{equation}
d_z = -e \langle \widetilde{\psi} | r_z | \widetilde{\psi} \rangle
= -2 e \sum_{| k_2 \rangle} \frac{\langle k_1 | r_z | k_2 \rangle
\langle k_2 | U_{\rm oct} | k_1 \rangle}{E_{k_1} - E_{k_2}},
\label{eaedm3a}
\end{equation}
where $\widetilde{\psi}$ denotes the perturbed atomic wave function,
$| k_1 \rangle = |n_1,j_1,l_1,m \rangle$ is the unperturbed
single-electron ground state,
and $\{| k_2 \rangle \}$ is the set of states with
which $| k_1 \rangle$ is mixed by the perturbation.

According to the triangle rule for the addition of angular
momenta, $\langle k_1 | r_z | k_2 \rangle$ can only have a nonzero
value if $|j_1 - j_2| \le 1 \le j_1 + j_2$. Similarly, for
$\langle k_2 | U_{\rm oct} | k_1 \rangle$ to be nonzero,
we must have
$|j_1 - j_2| \le 3 \le j_1 + j_2$. This implies
that the following conditions need to be satisfied
for the dipole moment to be nonzero:
\begin{equation}
|j_1 - j_2| \le 1 \mbox{ and } j_1 + j_2 \ge 3.
\label{eaedm3b}
\end{equation}
The lowest pair of values that satisfies this condition
is $j_1 = 3/2$ and $j_2 = 3/2$.
Therefore, $s$ states cannot contribute
to the dipole moment induced by the nuclear EOM. Also,
one of $l_1$ and $l_2$ must be even and the other odd, since the
electric dipole moment is a parity nonconserving effect.

We will carry out a relativistic calculation of the matrix
element of $U_{\rm oct}$ between the single-electron states
$|n_1,j_1,l_1,m \rangle$ and $|n_2,j_2,l_2,m \rangle$.
(Note that although we wrote Eq.\ (\ref{eaedm3a}) for an atom with
one electron above closed subshells, this single-electron state
matrix element can be used for all atoms to compare various
sources of atomic EDMs.)
The relativistic wave function of an electron
is (see, e.g., \cite{RelQuant})
\begin{equation}
\psi_{njlm} =
\left(
\begin{array}{c}
f_{njl} (r) \Omega_{jlm} (\theta,\phi) \\
g_{njl} (r) i (-\bbox{\sigma} \cdot {\bf n})
\Omega_{jlm} (\theta,\phi)
\end{array}
\right),
\label{eedm5}
\end{equation}
where $\bbox{\sigma}$ is twice the spin operator of this electron
and ${\bf n} = {\bf r}/r$.
Evaluating the matrix element (using $(\bbox{\sigma}
\cdot {\bf n})^2 = 1$) gives
\begin{equation}
\langle n_1 j_1 l_1 m | U_{\rm oct} | n_2 j_2 l_2 m \rangle
= -5 \sqrt{\frac{\pi}{7}} e {\cal O}
\langle j_1 l_1 m | Y_{30} | j_2 l_2 m \rangle T,
\label{eedm6}
\end{equation}
where $T$ is a radial integral:
\begin{equation}
T = \int_0^{\infty} \frac{1}{r^4}
(f_{n_1 j_1 l_1} f_{n_2 j_2 l_2}
+ g_{n_1 j_1 l_1} g_{n_2 j_2 l_2}) r^2 \, dr.
\label{eedm7}
\end{equation}
Because of the factor of $1/r^4$ in the above integral,
most of the contribution to $T$ comes from small values
of $r$. This allows us to use the following expressions
for $f_{njl}$ and $g_{njl}$, for
$r \ll a/Z^{1/3}$ \cite{Khripl}:
\begin{eqnarray}
f_{njl} (r) & = & \frac{c_{njl}}{r}
[(\gamma+\kappa) J_{2\gamma}(x)
- \frac{x}{2} J_{2\gamma-1}(x)], \nonumber \\
g_{njl} (r) & = & \frac{c_{njl}}{r} Z
\alpha J_{2\gamma}(x),
\label{eedm8}
\end{eqnarray}
where
\begin{eqnarray}
x & = & \sqrt{\frac{8 Z r}{a}}, \nonumber \\
\gamma & = & \sqrt{(j+{\textstyle \frac{1}{2}})^2
- Z^2 \alpha^2}, \nonumber \\
\kappa & = & (-1)^{j+1/2-l} (j+{\textstyle
\frac{1}{2}}), \nonumber \\
c_{njl} & = & \frac{\kappa}{|\kappa|}
\left( \frac{1}{Z a \nu^3} \right)^{1/2},
\label{eedm9}
\end{eqnarray}     
where the $J$'s are Bessel functions, $a$ is the
Bohr radius, and $\nu$ is the effective principal quantum
number ($E_{nl} = -13.6 \mbox{ eV} / \nu^2$). 
To avoid confusion, note that $l$ here is the orbital
angular momentum of the electron, rather than the nucleus, as 
used in Sec. \ref{sneom}. Also, the $\kappa$ used
here is distinct from
the $\kappa$ defined in Eq.\ (\ref{eao}).
Carrying out the integration in Eq.\ (\ref{eedm7}), we obtain
\begin{eqnarray}
\label{eedm10}
T & = & \frac{192 (-1)^{j_2-j_1+1} Z^2}{a^4 \nu_{1}^{3/2}
\nu_{2}^{3/2}} \frac{\Gamma (-3+\gamma_1
+\gamma_2)}{\Gamma(4+\gamma_1+\gamma_2)
\Gamma(4+\gamma_1-\gamma_2) \Gamma(4-\gamma_1+\gamma_2)}
\\ \nonumber
 & \times & \{(3+\gamma_1-\gamma_2) (3-\gamma_1+\gamma_2)
(3+\gamma_1+\gamma_2) (2+\gamma_1+\gamma_2) \\ \nonumber
&&-5 (\gamma_1+\kappa_1)(3-\gamma_1+\gamma_2)(3+\gamma_1+\gamma_2)
-5 (\gamma_2+\kappa_2)(3+\gamma_1-\gamma_2)(3+\gamma_1+\gamma_2) \\
\nonumber
&&+30 [(\gamma_1+\kappa_1)(\gamma_2+\kappa_2)+(Z \alpha)^2] \},
\end{eqnarray}
where $\Gamma$ is the gamma function.

To illustrate the numerical values involved, we will
evaluate a value
of the matrix element for an electron's interaction with the
single particle octupole moment for $^{209}$Bi.
This atom has one $6 p_{3/2}$ electron above closed subshells
($(6 p_{1/2})^2 (6 s_{1/2})^2 \ldots$). 
We will consider
mixing between $d_{5/2}$ and $p_{3/2}$ states, each with
an angular momentum projection of $3/2$. Using
table \ref{tsingp} and Eqs.\ (\ref{eedm6}) and (\ref{eedm10}) gives
\begin{equation}
\langle n_1 p_{3/2},m=3/2 | U_{\rm oct} | n_2 d_{5/2},m=3/2 \rangle
\approx 1.9 \times
10^{-13} (\nu_1 \nu_2)^{-3/2} \eta {\rm cm}^{-1}.
\label{eaedmme1}
\end{equation}
We now compare this with the corresponding matrix element for the
interaction with a nuclear magnetic quadrupole
moment (MQM), which is
another possible source of an atomic EDM. The MQM induced by the T-
and P-odd interaction (\ref{ehpta}) was calculated
in Ref. \cite{SFK84},
as well as the matrix element of the interaction between
an electron
and the MQM field.
For $^{209}$Bi the MQM is $\approx 4.8 \times 10^{-8} \eta e
(\mbox{fm})^2$. 
We get the following estimate:
\begin{equation}
\langle n_1 p_{3/2},m=3/2 | U_{\rm quad} | n_2
d_{5/2},m=3/2 \rangle
\approx 1.3 \times 10^{-11} (\nu_1 \nu_2)^{-3/2}
\eta {\rm cm}^{-1}.
\label{eaedmme2}
\end{equation}
By comparing the values of these matrix elements it
can be seen that the atomic EDM
induced by a nuclear EOM will be less than about 1\% of
that induced by a nuclear magnetic quadrupole moment.
This same result
held for the other atoms and mixing states that we considered.
Another mechanism which could produce an atomic EDM is the nuclear
Schiff moment, which for heavy atoms (which we are
considering) gives a
contribution comparable to that of
the MQM \cite{SFK84}. Because
of this we can conclude that the atomic EDM induced by a
single particle nuclear EOM
is negligible in comparison with other possible mechanisms.
(See the appendix for a discussion of the relative contributions
of the octupole, magnetic quadrupole, and Schiff moments.)

Now we will give example values of the matrix elements for those
nuclei having a static octupole deformation.
As well as possibly having collective EOMs (as shown in Sec.
\ref{sceom}), these nuclei can also have collective MQMs that
are an order of magnitude larger than the single particle MQMs
discussed above \cite{spinhh}. We will use $^{225}$Ac as an example
and once again we will consider mixing between $d_{5/2}$ and
$p_{3/2}$ states, each having an angular momentum projection of
$3/2$. This nucleus has an MQM of $\sim 2 \times 10^{-7} \eta e
(\mbox{fm})^2$ \cite{spinhh}.  
We obtain
\begin{equation}
|\langle n_1 d_{5/2},m=3/2 | U_{\rm oct} | n_2 p_{3/2},m=3/2
\rangle|
\approx 1.4 \times 10^{-11} (\nu_1 \nu_2)^{-3/2} \eta {\rm cm}^{-1}
\label{ecme1}
\end{equation}
and
\begin{equation}
|\langle n_1 d_{5/2},m=3/2 | U_{\rm quad} | n_2
p_{3/2},m=3/2 \rangle|
\sim 6 \times 10^{-11} (\nu_1 \nu_2)^{-3/2} \eta {\rm cm}^{-1}.
\label{ecme2}
\end{equation}
These results show that the contribution of the collective
nuclear EOM to
the atomic EDM is smaller than that of the collective MQM.
This also applies to the other isotopes shown in table \ref{tcoll},
except for $^{229}$Pa. ($^{223}$Rn does have a fairly
large collective
octupole moment, but since it has a closed electron shell
the EOM does not contribute to the
atomic EDM.) For $^{229}$Pa the contribution of the
EOM may be comparable to that of the MQM, but as stated above, it
is not certain that this nucleus has a static octupole deformation.
 
\section{Possibly enhanced magnetic quadrupole moments in nuclei
with an octupole deformation}

Finally, we discuss a mechanism by which single particle
magnetic quadrupole moments could be enhanced in even-odd nuclei
with an octupole deformation. We will first consider the MQM of
such a nucleus in its intrinsic (body-fixed) frame and then
transform the MQM into its laboratory frame.

As in Sec. \ref{sceom} the wave function of the external nucleon in
the intrinsic frame is $\psi_{+K} ({\bf r}^{\prime})$, defined
by Eqs. (\ref{eae11a}), (\ref{eae11b}), and (\ref{eae11c}). The
MQM in the intrinsic frame is then
\begin{eqnarray}
M_{\rm intr} & = & \langle \psi_{+K} | \hat{M}_{z^{\prime}
z^{\prime}} | \psi_{+K} \rangle
\label{emqn1a} \\
& = & 2 \gamma \langle \psi_{1,+K} | \hat{M}_{z^{\prime}
z^{\prime}} | \psi_{2,+K} \rangle,
\label{emqn1b}
\end{eqnarray}
where $\hat{M}$ is the operator for the MQM (see Ref. \cite{SFK84}).
Note that the MQM is defined for the maximum value of the
projection of the angular momentum onto the $z^{\prime}$-axis,
so $K=j=I$ (for the ground state of the rotational band $I=j$).
Using a result from Ref. \cite{SFK84} we have
\begin{equation}
M_{\rm intr} = \gamma \frac{2I-1}{I+1} \frac{e}{m}
(\mu - q) \int R_1 R_2 r \, r^2 dr,
\label{emqc}
\end{equation}
where $\mu$ is the magnetic moment of the external nucleon in
nuclear magnetons and $q = 0$ ($1$) for a neutron (proton).
It should be noted that this (intrinsic frame)
single particle MQM differs from
the spherical nucleus 
single particle EOMs and MQMs (considered in Sec. \ref{sneom}
and Ref. \cite{SFK84}, respectively) in that the former is
generated due to the interaction $V_3$, coming from the nucleus's
octupole deformed shape [see Eqs. (\ref{eae11b})
and (\ref{eae21a})], while the latter is due to the interaction
$H_{TP}$ (\ref{ehpta}). The $H_{TP}$ interaction will
come into the current
situation when we transform into the lab. frame.

Since we are only working out a rough estimate of the MQM we take
the integral in the above equation to be
\begin{equation}
\int R_1 R_2 r \, r^2 dr \sim \frac{1}{2} r_0 A^{1/3},
\label{eqmd}
\end{equation}
where $r_0 = 1.1 \mbox{ fm}$. For both types of nucleons we
have $|\mu - q| \approx 1.8$ and so we can write
\begin{equation}
| M_{\rm intr}| \sim 0.2 A^{1/3} \frac{2I-1}{I+1} |\gamma| e
(\mbox{fm})^2.
\label{emqe}
\end{equation}

Now we turn to the MQM in the laboratory frame. The wave function
of the nucleus in the lab. frame is (as in Sec. \ref{sceom})
$\psi = \psi^+ + \alpha \psi^-$, where $\psi^{\pm}$ is defined
by Eqs. (\ref{eceom1a}) and (\ref{ecn21a}) and $\alpha$ is
given by Eq. (\ref{enad32}) --- this is where the interaction
$H_{TP}$ comes into the present situation.
The MQM in the lab. frame is then
\begin{eqnarray}
M_{\rm lab} & = & \langle \psi | \hat{M}_{zz} | \psi
\rangle \nonumber \\
& = & 2 \alpha \langle \psi^+ | \hat{M}_{zz} | \psi^-
\rangle \nonumber \\
& = & \alpha ( \langle I M K | \hat{M}_{zz} | I M K \rangle
- \langle I M -K | \hat{M}_{zz} | I M -K \rangle) \nonumber \\
& = & 2 \alpha \langle I M K | \hat{M}_{zz} | I M K \rangle.
\label{emqea2}
\end{eqnarray}
(Once again, the MQM is defined for $M$, the projection of the
angular momentum onto the $z$-axis, equal to $I$.)
Note that $\langle I M -K | \hat{M}_{zz} | I M -K \rangle
= - \langle I M K | \hat{M}_{zz} | I M K \rangle$, as
$\psi_{\pm K} = \psi_{1, \pm K} \pm \gamma
\psi_{2, \pm K}$ and so
$\langle \psi_{-K} | \hat{M}_{z^{\prime} z^{\prime}}
| \psi_{-K} \rangle = -\langle \psi_{+K} |
\hat{M}_{z^{\prime} z^{\prime}}| \psi_{+K} \rangle$. 

Now consider $\langle I M K | \hat{M}_{zz} | I M K \rangle$.
We have [from Eq. (\ref{ecn21a})]
\begin{equation}
\langle I M K  | \hat{M}_{zz} | I M K \rangle
= \frac{2I+1}{4 \pi} \int {D^{I *}_{M K}} (\phi,
\theta,0) \psi_K^{*} ({\bf r}^{\prime}) \hat{M}_{zz}
(\theta,\phi,0) \psi_K ({\bf r}^{\prime})
D^I_{M K} (\phi,\theta,0) \, d^3 r^{\prime} \, d \Omega.
\label{emqf}
\end{equation}
Transforming $\hat{M}_{zz}$ from the lab. frame to the intrinsic
($x^{\prime},y^{\prime},z^{\prime}$) frame gives
$\hat{M}_{zz} (\theta,\phi,0) = {D^{2 *}_{0 0}} (\phi,
\theta,0) \hat{M}_{z^{\prime} z^{\prime}} (\theta^{\prime},
\phi^{\prime},0)$ (see, e.g., \cite{Varshalovich}). Substituting
this into Eq. (\ref{emqf}) and using Eq. (\ref{emqn1a}) gives
(using a formula for the integral of $D$-functions from
\cite{Varshalovich})
\begin{eqnarray}
\langle I M K | \hat{M}_{zz} | I M K \rangle & = &
\frac{2I+1}{4 \pi} M_{\rm intr} \int {D^{I *}_{M K}} (\phi,\theta,
0) {D^{2 *}_{0 0}} (\phi,\theta,0) D^I_{M K} (\phi,\theta,0)
\, d \Omega \nonumber \\
& = & M_{\rm intr} \langle I, M ; 2, 0|I, M \rangle
\langle I, K; 2, 0| I, K \rangle \nonumber \\
& = & M_{\rm intr} {\langle I, I ; 2, 0| I, I \rangle}^2
\nonumber \\
& = & \frac{I (2I-1)}{(I+1) (2I+3)} M_{\rm intr}.
\label{emqg}
\end{eqnarray}
(Recall that $M=K=I$.) Using Eq. (\ref{emqea2}) then gives
\begin{equation}
M_{\rm lab} = 2 \alpha \frac{I (2I-1)}{(I+1) (2I+3)}
M_{\rm intr}
\label{emq5a}
\end{equation}
and so [using Eq. (\ref{emqe})]
\begin{equation}
|M_{\rm lab}| \sim 0.4 A^{1/3} |\alpha| |\gamma|
\frac{I (2I-1)^2}{(I+1)^2 (2I+3)} e ({\rm fm})^2.
\label{emq5b}
\end{equation}
Now we have $|\alpha| \sim 7 \times 10^{-7} \eta$ and
$|\gamma| \sim 0.1$ (see Sec. \ref{sceom}). For
$A \approx 225$ we have
\begin{equation}
|M_{\rm lab}| \sim 6 \times 10^{-8} \eta e ({\rm fm})^2.
\label{emq5c}
\end{equation}
This is smaller than the collective MQM due to the spin
hedgehog mechanism ---
$\sim 2 \times 10^{-7} \eta e ({\rm fm})^2$ \cite{spinhh} and
so it is not the dominant mechanism. It is, in
fact, of the same order of magnitude as the single particle
MQM [e.g., for $^{209}$Bi, approximately
$4.8 \times 10^{-8} \eta e ({\rm fm})^2$ (see
Sec. \ref{saedmi})].
This means that there is no enhancement.

Enhancement was a possibility here due to the relatively large
value of $\alpha$ that comes from the smallness of the energy
splitting between members of the parity doublet ($E_+ - E_-$).
However, the inclusion of the factors of $|\gamma| \sim 0.1$
and $I (2I-1) / [(I+1) (2I+3)] \sim 0.3$ (this enters on
transforming to the lab. frame) ensures that the possible
enhancement is not realised.

However, $^{229}$Pa may be an exception to this, as it has a
large value of $\alpha$, but it must be remembered that
it is not certain that this nucleus has a static octupole
deformation (see Sec. \ref{sceom}). If is does have such a
deformation then its MQM due to the present mechanism would
be $|M_{\rm lab}| \sim 3 \times 10^{-7}
\eta e ({\rm fm})^2$ [using Eq. (\ref{emq5b}),
the value of $\alpha$ given in Ref. \cite{AFS2}, and
$|\gamma| \sim 0.1$], which is of the same order of
magnitude as the collective MQM due to the spin hedgehog
mechanism.

Note that the enhancement of MQMs in
deformed nuclei with opposite
parity levels close to each other has also been considered in
Refs. \cite{Hax83,SFK84}.

\begin{acknowledgements}
One of us (DWM) is grateful to G.F. Gribakin for
helpful discussions.
This work was supported by the Australian Research Council
and by the National Science Foundation through a grant for the
Institute for Theoretical Atomic and Molecular Physics at
Harvard University and the Smithsonian Astrophysical Observatory.
\end{acknowledgements}

\appendix
\section{The relative contributions of various
nuclear moments to the atomic electric dipole moment}
\label{sappendix}
An atomic electric dipole moment can be generated by T-,P-odd
nuclear moments. However, according to the
Purcell-Ramsey-Schiff theorem \cite{Purcell50,Schiff63}, a
nuclear EDM cannot generate an atomic EDM. This happens
because the electrostatic potential of a
nucleus is screened by its atomic electrons.
Therefore we must look to higher order moments.

The (screened) electrostatic
potential of the nucleus can be written as
(see, e.g., \cite{SFK84,AFS96,AFS2})
\begin{equation}
\phi ({\bf R}) = \int \frac{e \rho({\bf r})}{|{\bf R}-{\bf r}|}
\, d^3 r + \frac{1}{Z} ({\bf d} \cdot \bbox{\nabla})
\int \frac{\rho({\bf r})} {|{\bf R}-{\bf r}|} \, d^3 r,
\label{eia1a}
\end{equation}
where $\rho({\bf r})$ is the nuclear charge density
($\int \rho({\bf r}) \, d^3 r = Z$),
${\bf d} = \int e {\bf r} \rho({\bf r}) \, d^3 r$
is the nuclear EDM, and $\nabla_i \equiv
\partial_i \equiv \frac{\partial}{\partial R_i}$.
The first non-zero T-,P-odd term in this
potential is \cite{SFK84,AFS96,AFS2}
\begin{eqnarray}
\phi^{(3)} & = & -\frac{1}{6} \int e \rho({\bf r}) r_i r_j r_k
\, d^3 r \, \partial_i \partial_j \partial_k \frac{1}{R}
+ \frac{1}{2Z} \int \rho({\bf r}) r_i r_j \, d^3 r \,
({\bf d} \cdot \bbox{\nabla}) \partial_i
\partial_j \frac{1}{R} \nonumber \\
& = & -\frac{1}{6} \int e \rho({\bf r}) [r_i r_j r_k
-\frac{1}{5} r^2 (r_i \delta_{jk} + r_j \delta_{ik}
+ r_k \delta_{ij})] \,
d^3 r \, \partial_i \partial_j \partial_k \frac{1}{R}
\nonumber \\
& & -\frac{1}{30} \int e \rho({\bf r}) 
r^2 (r_i \delta_{jk} + r_j \delta_{ik}
+ r_k \delta_{ij})
\, d^3 r \, \partial_i \partial_j \partial_k \frac{1}{R}
\nonumber \\
& & + \frac{1}{2Z} \int \rho({\bf r}) (r_i r_j
-\frac{1}{3} r^2 \delta_{ij}) \, d^3 r \, ({\bf d} \cdot
\bbox{\nabla}) \partial_i \partial_j \frac{1}{R}
\nonumber \\
& & + \frac{1}{6Z} \int \rho({\bf r})
r^2 \delta_{ij}
\, d^3 r \, ({\bf d} \cdot \bbox{\nabla}) \partial_i
\partial_j \frac{1}{R} \nonumber \\
& = & -\frac{1}{6} O_{ijk} \partial_i \partial_j
\partial_k \frac{1}{R}
+ \frac{1}{e} \frac{1}{2 Z} Q_{ij} \, ({\bf d} \cdot \bbox{\nabla})
\partial_i \partial_j \frac{1}{R} \nonumber \\ 
& & - \frac{1}{10} \int e \rho({\bf r}) r^2 r_i \, d^3 r \,
\partial_i \delta_{jk} \partial_j \partial_k \frac{1}{R}
+ \frac{1}{6 Z} \int \rho({\bf r}) r^2 \, d^3 r \,
({\bf d} \cdot \bbox{\nabla}) \delta_{ij} \partial_i
\partial_j \frac{1}{R} \nonumber \\
& = & \phi^{(3)}_{\mbox{octupole}} + \phi^{(3)}_{\mbox{Schiff}},    
\label{eia2a}
\end{eqnarray}
where
\begin{eqnarray}
\phi^{(3)}_{\mbox{octupole}} & = & -\frac{1}{6} O_{ijk}
\partial_i \partial_j \partial_k \frac{1}{R}
+ \frac{1}{e} \frac{1}{2Z} Q_{ij} d_k
\partial_i \partial_j \partial_k \frac{1}{R} \nonumber \\
& \approx & -\frac{1}{6} O_{ijk}
\partial_i \partial_j \partial_k \frac{1}{R},
\label{een08baa} \\
\phi^{(3)}_{\mbox{Schiff}} & = &
-{\bf S} \cdot \bbox{\nabla}
\nabla^2 \frac{1}{R} = 4 \pi {\bf S} \cdot
\bbox{\nabla} \delta(R)
\label{een08b}
\end{eqnarray}
(using $\delta_{jk} \partial_j \partial_k R^{-1}
= \nabla^2 R^{-1} = -4 \pi \delta(R)$).
$Q_{ij}$ is the T,P-even electric quadrupole moment:
$Q_{ij} = \int e \rho({\bf r}) (r_i r_j - \frac{1}{3} r^2
\delta_{ij}) \, d^3 r$.
The vector ${\bf S}$ is the nuclear Schiff moment and the rank
$3$ tensor $O_{ijk}$ is the nuclear electric octupole moment. These
are given by
\begin{eqnarray}
O_{ijk} & = & \int e \rho({\bf r}) [r_i r_j r_k
- \frac{1}{5} r^2 (r_i \delta_{jk} + r_j \delta_{ik}
+ r_k \delta_{ij})] \, d^3 r,
\label{eneo8caa} \\
{\bf S} & = & \frac{1}{10} \left(
\int e \rho({\bf r}) r^2 {\bf r} \, d^3 r
-\frac{5}{3} {\bf d} \frac{1}{Z}
\int \rho({\bf r}) r^2 \, d^3 r \right).
\label{ene08c}
\end{eqnarray}

The second term in the octupole potential (\ref{een08baa})
comes from the screening of the nucleus's Coulomb field.
Because only the non-spherically symmetric part of the density
will give a non-zero value of $Q_{ij}$, only the external
proton contributes to this screening term (in the case of
a deformed nucleus all the protons in the external shell
contribute). This, together with the presence of the
factor $1/Z$, means that it will be small and so
we neglect it. The screening term in the Schiff moment
[the second term of Eq. (\ref{ene08c})] is not negligible
as the whole density $\rho({\bf r})$ contributes to it
and so all of the protons are involved.

$\phi^{(3)}_{\mbox{Schiff}}$ and $\phi^{(3)}_{\mbox{octupole}}$
both appear at the same order in the expansion of
Eq. (\ref{eia1a}), i.e., both potentials
contain third derivatives of $1/R$
and their corresponding nuclear moments are
both integrals with integrands containing $r^3$ terms.
Therefore, we should expect the contributions of
the Schiff and octupole moments to the atomic EDM to be
roughly the same. However, since the octupole moment is a rank
$3$ tensor, as opposed to the (vector) Schiff moment, an
electron interacting with an octupole moment must have a
higher angular momentum (see Sec. \ref{saedmi}) than it would need
for a Schiff moment. This results in the octupole moment's
contribution to the atomic EDM being smaller than the
Schiff moment's, as a higher angular momentum wave function
penetrates the region close to the nucleus less due to the
greater centrifugal barrier.
This is confirmed by the calculations in
this paper (see Sec. \ref{saedmi}).

An atomic EDM can also be caused by a nuclear magnetic quadrupole
moment (MQM). The vector potential due to such
a MQM is \cite{SFK84}
\begin{equation}
A_{i} ({\bf R}) = - \frac{1}{6} \varepsilon_{iln}
M_{kn} \partial_l \partial_k \frac{1}{R},
\label{eia4a}
\end{equation}
where $M_{kn}$ is the (rank $2$) magnetic quadrupole moment
tensor:
\begin{equation}
M_{kn} = - \int (r_k \varepsilon_{npq} + r_n \varepsilon_{kpq})
j_p r_q \, d^3 r
\label{eia4b}
\end{equation}
($j_p$ is the electromagnetic current.) Note that
this is a lower
order term than the Schiff and octupole moments --- the potential
contains second derivatives of $1/R$
and the integrand in Eq. (\ref{eia4b})
contains $r^2$ terms.
However, the contribution of the MQM to the atomic EDM will
actually be of roughly the same order of magnitude as that of
the Schiff and octupole moments. To see this, consider the
(nonrelativistic) interaction between the atomic electron's
magnetic dipole moment and the magnetic field from the MQM:
$\mu_e B = \mu_e (\bbox{\nabla} \times {\bf A})$, where
$\mu_e$ is the Bohr magneton. Now from Eq. (\ref{eia4a})
we have $\bbox{\nabla} \times {\bf A} \sim \bbox{\nabla}
\times (M / R^3) \sim M / R^4$, where $M$ is the magnetic
quadrupole moment. From Ref. \cite{SFK84} we have
$M \sim \mu_N \xi$, where $\mu_N$ is the nuclear magneton and
$\xi$ is defined by Eq. (\ref{ehptc}). Therefore we have
(in $\hbar = c = 1$ units)
\begin{equation}
\mu_e B \sim \frac{\mu_e \mu_N}{R^4} \xi
\sim \frac{e}{m_e} \frac{e}{m_p} \frac{1}{R^4} \xi
\label{eea07a}
\end{equation}
Now we compare this with the electron's interaction with the
octupole:
\begin{equation}
e \phi^{(3)}_{\mbox{octupole}} \sim \frac{e {\cal O}}{R^4}
\sim \frac{e^2 {R_N}^2}{R^4} \xi,
\label{eea07b}
\end{equation}
using the result for the octupole moment in Eq. (\ref{eau})
($R_N = 1.1 A^{1/3} \mbox{ fm}$ is the radius of the nucleus).
Therefore, the ratio of the contribution of the octupole moment
to the atomic EDM to the contribution of the MQM is
$\sim {R_N}^2 m_e m_p \sim 10^{-2} A^{2/3}$.
A more accurate estimate that takes into account
the electron angular momenta dependence of the matrix
elements gives a ratio that is closer to $10^{-4} A^{2/3}$.
In Sec. \ref{saedmi} the ratio was
$\sim 0.01$, which is consistent with this result.

\begin{table}
\begin{center}
\begin{tabular}{l c r}
Nucleus & Proton state & ${\cal O}_{\rm sing}$ [$\eta e
({\rm fm})^3$] \\
\hline
$^{209}$Bi & $h_{9/2}$ & $1.2 \times 10^{-7}$ \\
$^{133}$Cs & $g_{7/2}$ & $7 \times 10^{-8}$ \\
$^{127}$I & $d_{5/2}$ & $-1.3 \times 10^{-7}$
\end{tabular}
\end{center}
\caption{Approximate values of the unpaired proton octupole moment
(${\cal O}_{\rm sing}$) for various nuclei with $I \ge 3/2$,
in terms of the parameter $\eta$.}
\label{tsingp}
\end{table}
 
\begin{table}
\begin{center}
\begin{tabular}{l c r}
Nucleus & $I$ & 
$|{\cal O}_{\rm coll}|$ [$\eta e ({\rm fm})^3$] \\ \hline
$^{223}$Ra & $\frac{3}{2}$ & $2 \times 10^{-6}$ \\
$^{223}$Rn & $\frac{7}{2}$ & $7 \times 10^{-5}$ \\
$^{225}$Ac & $\frac{3}{2}$ & $7 \times 10^{-6}$ \\
$^{223}$Fr & $\frac{3}{2}$ & $4 \times 10^{-6}$ \\
$^{229}$Pa & $\frac{5}{2}$ & $3 \times 10^{-4}$
\end{tabular}
\end{center}
\caption{Approximate values of the collective octupole moment
(${\cal O}_{\rm coll}$) for various nuclei that have a
static octupole deformation and $I \ge 3/2$, in terms of the
parameter $\eta$.}
\label{tcoll}
\end{table}
\end{document}